\documentclass[letterpaper, 10 pt, conference]{ieeeconf}  

\IEEEoverridecommandlockouts                              

\usepackage{cite}
\usepackage{amsmath,amssymb,amsfonts}
\usepackage{graphicx}
\usepackage{textcomp}
\usepackage{xcolor}
\usepackage{hyperref}
\usepackage{algorithm}
\usepackage{algpseudocode}
\usepackage{soul}
\usepackage{subcaption}
\def\BibTeX{{\rm B\kern-.05em{\sc i\kern-.025em b}\kern-.08em
    T\kern-.1667em\lower.7ex\hbox{E}\kern-.125emX}}

\usepackage{tikz}
\usepackage{enumerate}

\graphicspath{{./Pictures/}}

\newtheorem{theorem}{Theorem}
\newtheorem{example}{Example}
\newtheorem{remark}{Remark}

\newtheorem{lemma}{Lemma}

\newcommand{\rs}[1]{r_{s_{#1}}}
\newcommand{\rc}[1]{r_{c_{#1}}}
\newcommand{\xs}[1]{x_{#1}}
\newcommand{\ps}[1]{p_{s_{#1}}}
\newcommand{\pc}[1]{p_{c_{#1}}}

\newcommand{\F}{\mathcal{F}}
\newcommand{\N}{\mathcal{N}}
\newcommand{\Nn}{\mathbb{N}}

\newcommand{\obs}{\mathcal{O}}

\title{\LARGE \bf A Distributed Gradient-Based Deployment Strategy for a Network of Probabilistic Sensors}

\author{Hesam Mosalli
and Amir G. Aghdam
\thanks{
This work has been supported by the Natural Sciences and
Engineering Research Council of Canada (NSERC) under grant RGPIN-2024-06367, and by the Fonds de recherche du Québec -- Nature et technologies (FRQNT) under funding \href{https://doi.org/10.69777/344572}{DOI: 10.69777/344572}.}
\thanks{H. Mosalli and A. G. Aghdam are with the Department of
Electrical and Computer Engineering, Concordia University, Montréal, QC, Canada. Email: \href{mailto:hesam.mosalli@mail.concordia.ca}{hesam.mosalli@mail.concordia.ca}, \href{mailto:amir.aghdam@concordia.ca}{amir.aghdam@concordia.ca}}
\thanks{The authors acknowledge the use of Grammarly and AI agents in assisting with the writing refinement of this manuscript.}
}

\begin{document}

\maketitle

\begin{abstract}
This paper presents a distributed gradient-based deployment strategy to maximize coverage in hybrid wireless sensor networks (WSNs) with probabilistic sensing. Leveraging Voronoi partitioning, the overall coverage is reformulated as a sum of local contributions, enabling mobile sensors to optimize their positions using only local information. The strategy adopts the Elfes model to capture detection uncertainty and introduces a dynamic step size based on the gradient of the local coverage, ensuring movements adaptive to regional importance. Obstacle awareness is integrated via visibility constraints, projecting sensor positions to unobstructed paths. A threshold-based decision rule ensures movement occurs only for sufficiently large coverage gains, with convergence achieved when all sensors and their neighbors stop at a local maximum configuration. Simulations demonstrate improved coverage over static deployments, highlighting scalability and practicality for real-world applications.
\end{abstract}

\section{Introduction}
\label{sec: intro}
Wireless sensor networks (WSNs) have revolutionized real-time monitoring, enabling applications from environmental observation to disaster management and military surveillance~\cite{akyildiz2002wireless, lin, mahamuni2016military}. Comprising low-cost, battery-powered nodes, WSNs are deployed across a region of interest (ROI) to gather data on parameters such as temperature, humidity, or motion~\cite{abbasi2007survey}. The effectiveness of these networks hinges on coverage—the extent to which the ROI is monitored, signifying optimized sensor placement and movement. Yet, real-world deployments face multifaceted challenges: sensor heterogeneity, probabilistic sensing, physical obstacles, energy constraints, and dynamic environments complicate traditional optimization strategies, necessitating innovative approaches~\cite{zou2003, wang2013, elhabyan2019coverage}.

Coverage optimization seeks to maximize the monitored area while adhering to practical limitations like energy consumption, communication bandwidth, and scalability~\cite{mahboubi2017}. In static WSNs, sensors remain fixed after deployment, offering stability but lacking adaptability. In contrast, mobile sensor networks or hybrid networks with both stationary and mobile nodes can reposition to enhance coverage or respond to changes, though this introduces energy and coordination overheads~\cite{cortez2004, mosalli2023}. Heterogeneity—where sensors vary in sensing range, energy capacity, or reliability—introduces further challenges. For instance, a long-range temperature sensor contributes differently to coverage than a short-range motion detector, requiring strategies that account for individual capabilities~\cite{mahboubi2013, habibi2017}. Probabilistic sensing models add another layer of complexity, as detection probability decreases with distance or environmental noise, deviating from binary ideals. The Elfes model, for example, captures this decay, offering a realistic alternative to deterministic assumptions~\cite{elfes1989using, hefeeda2007probabilistic}. Physical obstacles—such as trees, buildings, or terrain—block sensing and communication, creating coverage gaps, while energy constraints limit mobile node repositioning, risking network degradation over time~\cite{wang2013, yoon2011coordinated}.

Coverage optimization in WSNs has been extensively studied in the literature. Early research focused on homogeneous networks, employing methods like Voronoi diagrams or grid-based positioning to enhance coverage with minimal sensors~\cite{okabe,zou2003}. While useful under ideal conditions, these approaches are not as effective in more complex settings~\cite{dhillon2003sensor}. Gradient-based methods have since emerged as a promising alternative for distributed optimization. For instance,~\cite{habibi2017} adapts this technique to heterogeneous WSNs, optimizing deployment for nonidentical sensing ranges, and~\cite{cortez2004} applies it to mobile networks~\cite{mahboubi2014}. Probabilistic sensing has been considered in~\cite{hefeeda2007probabilistic} to optimize expected coverage under uncertainty, and in~\cite{jia2009energy}, to balance coverage with energy efficiency. Hybrid WSNs, combining mobile and stationary nodes, have been investigated in~\cite{mosalli2023}, proposing distributed strategies for both node types, and in~\cite{guo}, focusing on cooperative coverage.

The presence of obstacles in real-world environments further complexifies the coverage problem. Virtual force algorithms, as in~\cite{mahfoudh2013relocation}, adjust sensor positions by repelling them from barriers and attracting them to uncovered areas. The authors in~\cite{wang2013} use sensor data to detect and adapt to obstacles. Energy efficiency, on the other hand, is a critical concern that has driven innovations like coordinated locomotion in~\cite{yoon2011coordinated} and potential field-based control in~\cite{kwok2011}. Distributed control, emphasized in~\cite{cortez2004,hussain2024comprehensive}, leverages local information for scalability. Probabilistic models, such as the Elfes model in~\cite{elfes1989using,elhabyan2019coverage}, redefine coverage as an expected value, challenging binary assumptions. Machine learning approaches, like reinforcement learning in~\cite{karegar2024deep,mosalli2024}, offer adaptability but are computationally intensive~\cite{feriani2021}, while dynamic coverage control in~\cite{yoon2011coordinated} addresses time-varying environments. However, integrating heterogeneity, probabilistic sensing, obstacles, and energy constraints remains an open challenge~\cite{rawat2024energy,abbasi2007survey,wang2011coverage}.

The primary contribution of this paper is a distributed gradient-based deployment strategy tailored for hybrid homogeneous WSNs with probabilistic sensing. Building on~\cite{habibi2017,mosalli2023}, it reformulates coverage as a sum of local contributions using Voronoi partitioning, incorporates the Elfes model~\cite{elfes1989using} for realistic detection uncertainty, and introduces a dynamic step size to balance adaptability and energy efficiency. Unlike prior work, it considers both probabilistic sensing and obstacle awareness in a scalable framework, ensuring mobile sensors optimize coverage locally while respecting visibility and energy constraints. This approach targets critical applications like disaster response and surveillance, offering a practical solution where coverage gaps are costly.

This paper is structured as follows. Section~\ref{sec: problem} defines the problem and provides the required background. Section~\ref{sec: dist_cov} reformulates coverage for distributed optimization. Section~\ref{sec: strategy} details the proposed strategy, and Section~\ref{sec: sim} presents simulation results. Finally, Section~\ref{sec: conc} concludes the paper and suggests directions for future research.

\section{Preliminaries and Problem Statement}
\label{sec: problem}

This research tackles coverage optimization in a wireless sensor network deployed over a 2D region of interest (ROI), denoted by \(\F \subset \mathbb{R}^2\), where monitoring demands vary across space. A priority function \(\varphi: \F \to \mathbb{R}^{\geq 0}\) assigns significance to points \(q \in \F\), with \(\varphi(q)\) indicating the importance of covering \(q\). Obstacles, represented by a set \(\mathcal{O}\), obstruct sensing but not communication and are excluded from \(\F\) to prioritize unobstructed regions. The WSN consists of \(n\) identical sensors \(\{s_1, s_2, \dots, s_n\}\), each positioned at \(\xs{i} \in \F\) with a sensing model \(\ps{i}\) and communication model \(\pc{i}\), for \(i \in \Nn_n := \{1, 2, \dots, n\}\). The network operates as a hybrid system: sensors \(s_1, \dots, s_m\) (\(m \leq n\)) are mobile and can reposition, while \(s_{m+1}, \dots, s_n\) are stationary with fixed locations. The goal here is to devise a gradient-based deployment strategy that repositions mobile sensors to maximize coverage, relying on local information in a distributed fashion.

\subsection{Sensing Model}
\begin{figure}[t]
\begin{subfigure}[t]{0.45\linewidth}
    \centering
    \begin{tikzpicture}[scale=0.4]
        \fill[blue!40] (0,0) circle (2);
        \draw[thick] (0,0) circle (2);
        \node at (0,0) [circle, draw=black, fill=black, inner sep=2pt, label=below:$\xs{i}$] {};
        \draw[-,thick, dashed] (0,0) -- (1.414,1.414) node[midway,left] {\(\rs{}\)};
    \end{tikzpicture}
    \caption*{(a)}
    \end{subfigure}
    \hfill
    \begin{subfigure}[t]{0.45\linewidth}
        \centering
    \begin{tikzpicture}[scale=1]
        \fill[inner color=blue!80, outer color=blue!10] (0,0) circle (2);
        \fill[blue!50] (0,0) circle (1);
        \draw[thick] (0,0) circle (1);
        \draw[thick] (0,0) circle (2);
        \node at (0,0) [circle, draw=black, fill=black, inner sep=2pt, label=below:$\xs{i}$] {};
        \draw[-,thick, dashed] (0,0) -- (0,1) node[midway,left] {\(\rs{}^{\min}\)};
        \draw[-,thick, dashed] (0,0) -- (1.414,1.414) node[midway,below right] {\(\rs{}^{\max}\)};
    \end{tikzpicture}
    \caption*{(b)}
    \end{subfigure}
    \caption{Comparison of sensing models: (a) deterministic model, (b) Elfes probabilistic model}
    \label{fig:sensingmodels}
\end{figure}
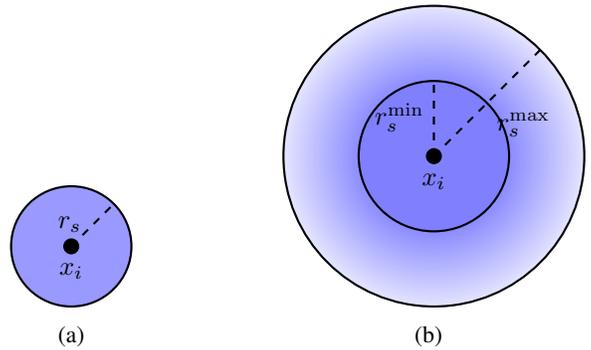

This work employs a probabilistic sensing model to reflect real-world detection uncertainty, adopting the Elfes model~\cite{elfes1989using}. For sensor \(s_i\), the sensing probability at point \(q \in \F\) is defined as:
\begin{equation}
\label{eq:elfes}
\ps{i}(q) = 
\begin{cases} 
1 & \text{if } \|\xs{i} - q\| \leq \rs{}^{\min}, \\
e^{-\alpha (\|\xs{i} - q\| - \rs{}^{\min})} & \text{if } \rs{}^{\min} < \|\xs{i} - q\| \leq \rs{}^{\max}, \\
0 & \text{otherwise},
\end{cases}
\end{equation}
where \(\|.\| \) denotes the Euclidean distance, \(\rs{}^{\min}\) is the radius of perfect detection, \(\rs{}^{\max}\) is the maximum sensing radius, and \(\alpha > 0\) specifies the decay rate of detection probability. Due to the assumption of homogeneity, \(\rs{}^{\min}\), \(\rs{}^{\max}\), and \(\alpha\) are the same for all sensors. In special case, setting \(\rs{}^{\min} = \rs{}^{\max} = \rs{}\) reduces the model to a deterministic one
with perfect detection within a fixed sensing radius \(\rs{}\) and no sensing beyond that. Figure~\ref{fig:sensingmodels} illustrates this difference, showing a sharp boundary for the deterministic model versus a gradual decay for the probabilistic Elfes model, as represented by gradually decreasing power density.

Sensing visibility becomes an essential consideration in the presence of obstacles. If obstacles block the line of sight from \(\xs{i}\) to \(q\), \(\ps{i}(q) = 0\). The sensing range of \(s_i\) is thus:
\begin{equation}
\label{eq:sensingrange}
D(s_i) = \{q \in \Phi(\xs{i}) \mid \ps{i}(q) > 0\},
\end{equation}
where \(\Phi(\xs{i})\) denotes the visible subset of \(\F\) from \(\xs{i}\).
\subsection{Communication Model}
In this work, the communication capability of sensors is modeled deterministically, described as:
\begin{equation}
\label{eq:comm}
\pc{i}(\xs{j}) = 
\begin{cases} 
1 & \text{if } \|\xs{i} - \xs{j}\| \leq \rc{}, \\
0 & \text{otherwise},
\end{cases}
\end{equation}
where \(\rc{}\) is the communication radius of each sensor. This assumes reliable communication within \(\rc{}\), unaffected by obstacles, offering a practical simplification for hybrid WSNs where connectivity is vital. In this model, \(s_j\) receives data from \(s_i\) if \(\pc{i}(\xs{j}) =1\). The communication structure can be represented by a graph \(\mathcal{G} = (\mathcal{V}, \mathcal{E})\), with \(\mathcal{V} = \Nn_n\) as nodes and $\mathcal{E}$ as edges, where the edge from $i$ to $j$, denoted by \((i, j)\) belongs to \(\mathcal{E}\) if $s_i$ and $s_j$ are connected. Since the sensors are identical, $\mathcal{G}$ is an undirected graph, and the sensor communications are bidirectional. Then, the set of neighbors of \(s_i\) is expressed as:
\begin{equation}
\label{eq:neighbors}
\mathcal{N}_i = \{j \in \Nn_n \mid j \neq i, \, \|\xs{i} - \xs{j}\| \leq \rc{}\},
\end{equation}
that is, the set of sensors with which \(s_i\) can exchange information. As mobile sensors reposition, \(\mathcal{G}\) evolves over time and needs to be updated at each iteration.

\subsection{Coverage Objective}
The network’s collective sensing function at \(q\) is given by:
\begin{equation}
\label{eq:collective}
\ps{}(q) = \max_{i \in \Nn_n} \ps{i}(q),
\end{equation}
using the maximum probability to represent non-overlapping contributions. The overall coverage is measured as:
\begin{equation}
\label{eq:coverage}
F = \int_{\F} \varphi(q) \ps{}(q) \, dq,
\end{equation}

The aim is to maximize \(F\) by repositioning mobile sensors while the stationary ones are fixed. The problem of finding the globally maximum coverage can be stated as follows: 
\begin{equation}
\label{eq: max_cov}
    \max_{\{\xs{i}\}_{i\in\Nn_m}}\quad F.
\end{equation}
However, this goal needs to be achieved by each sensor only using the local information obtained from its neighbors. Furthermore, instead of solving~\eqref{eq: max_cov} as a positioning problem, it is desired to guide the mobile sensors from an initial to a final configuration by a distributed deployment rule.

\section{Distributed Coverage Maximization Problem}
\label{sec: dist_cov}

This section presents a distributed gradient-based deployment strategy to maximize coverage in a hybrid WSN with probabilistic sensing. To address the problems, the overall coverage function is reformulated as the sum of local coverages, leveraging Voronoi partitioning to assign each sensor a region of responsibility. This approach enables distributed optimization, where mobile sensors adjust their positions based on local gradients, enhancing scalability and adaptability in the presence of obstacles.

The overall coverage \( F \), as defined in~\eqref{eq:coverage}, integrates the maximum sensing probability over the union of all sensors' sensing ranges. However, computing \( F \) directly requires global knowledge, which is impractical in large-scale WSNs. By partitioning the ROI \(\F\) into Voronoi regions, assigning each sensor \( s_i \) a region \(\pi_i\) containing all points \( q \in \F \) closer to \(\xs{i}\) than to any other sensor's position \(\xs{j}\) (\( j \neq i \)), it is possible to compute the overall coverage in a distributed manner. This can be mathematically described as:
\begin{equation}
\label{eq:voronoi}
\pi_i = \{ q \in \F \mid \|\xs{i} - q\| \leq \|\xs{j} - q\|, \forall j \in \Nn_n, j \neq i \}\, , i\in\Nn_n.
\end{equation}

For a homogeneous WSN with Elfes sensing model, the above inequality guarantees that each sensor has the highest sensing probability in its corresponding Voronoi region.

\begin{lemma}
\label{lemma: ps}
    Let the ROI be partitioned into disjoint Voronoi regions $\pi_1,\pi_2,\dots,\pi_n$ defined in~\eqref{eq:voronoi} with $n$ identical sensors as the generating nodes. Under Elfes sensing model, for every $q\in\pi_i$, $p_s(q)=\ps{i}(q)$.
\end{lemma}

\begin{proof}
By the definition of the Voronoi region $\pi_i$, any point $q\in\pi_i$ satisfies the following inequality:
\begin{equation*}
\|q-\xs{i}\| \;\le\; \|q-\xs{j}\| \quad \forall\, j\neq i,\quad j\in\Nn_n.
\end{equation*}
Since the Elfes model $\ps{i}(q)$ is monotonically decreasing with respect to $\|q-\xs{i}\|$ over the interval $[\rs{}^{\min},\;\rs{}^{\max}]$, having the smallest distance to $q$ implies the largest detection probability among all sensors. Thus,
\[
\ps{i}(q) \;\ge\; \ps{j}(q) \quad\text{for all } i\in \Nn_n, j\neq i,
\]
and therefore,
\[
p_s(q)\;=\;\max_{j\in\Nn_n}\,\ps{j}(q)\;=\;\ps{i}(q).
\]
This completes the proof.
\end{proof}

As a result of Lemma~\ref{lemma: ps}, the overall coverage can be approximated as the sum of local coverages over each sensor's Voronoi region, i.e.:
\begin{equation}
\label{eq:local_coverage}
F = \sum_{i \in \Nn_n} \int_{\pi_i} \varphi(q) \ps{}(q) \, dq.
\end{equation}
Since the sensing range of each sensor defined in~\eqref{eq:sensingrange} is limited by their maximum sensing radius and visibility, \eqref{eq:local_coverage} can be rewritten as:
\begin{equation}
F = \sum_{i \in \Nn_n}\int_{\pi_i \cap D(s_i)} \varphi(q) \ps{i}(q) \, dq.
\end{equation}

This reformulation assumes that each point \( q \) is primarily monitored by its nearest sensor, aligning with the aimed distributed control strategy for relocating the mobile sensors. However, due to the fixedness of the stationary sensors, this approach would fail to take the coverage holes inside their Voronoi regions into account. As a result, we will redefine the partitioning by considering only the mobile sensors as the generating nodes. Moreover, to seek a distributed solution, each sensor must be able to construct the Voronoi partitioning based on its neighbors' locations.

Suppose the sensing field is partitioned into a Voronoi diagram from a sensor $ s_i$'s perspective with itself and its mobile neighbors as the generating nodes and let $\Pi_i$ be the region corresponding to $s_i$. Then, the part of local coverage observable and assigned to $s_i$ is:
\begin{equation}
\label{eq:Fi}
F_i = \int_{\Pi^\prime_i \cap D(s_i)} \varphi(q) \ps{i}(q) \, dq,
\end{equation}
where $\Pi^\prime_i$ is part of $\Pi$ in which $\ps{i}\geq \ps{j}$ for all $j\in\N_i$. Note that by definition, this inequality is already satisfied for every mobile sensor and its mobile neighbors. Thus, to find $\Pi^\prime_i$, one should eliminate parts of the sensing ranges of those stationary neighbors that are closer to them than $s_i$. In other terms:
\begin{equation}
\label{eq:pip}
\begin{aligned}
    \Pi^\prime_i=\{q\in\Pi_i \mid ~ &\|\xs{i} - q\| \leq \|\xs{j} - q\| ~ \text{or} ~ q\notin D(s_j),\\
    &\forall j \in \N_i, j > m\}.
\end{aligned}
\end{equation}

In contrast to the Voronoi regions $\pi_1,\pi_2,\dots,\pi_n$ in~\eqref{eq:voronoi}, the regions $\Pi^\prime_1, \Pi^\prime_2,\dots,\Pi^\prime_m$ are not necessarily a disjoint partitioning of the ROI (nor are the areas to be covered by mobile sensors) because of the limited communication range of the sensors. However, it is still possible to restate the overall coverage in terms of the local coverages~\eqref{eq:Fi}, because the covered areas by the mobile sensors, that are $\Pi^\prime_i \cap D(s_i)$ for $i\in\Nn_m$, are disjoint.

\begin{lemma}
\label{lemma:fbreak}
    Consider the Voronoi partitioning of the ROI by $\pi_1,\pi_2,\dots,\pi_n$, and the regions $\Pi^\prime_1, \Pi^\prime_2,\dots,\Pi^\prime_m$ defined in~\eqref{eq:pip}. Then, the overall coverage $F$ can be written as:
    \begin{equation}
    \label{eq:fbreak}
    \begin{aligned}
        F=& \sum_{i=1}^m\int_{\Pi^\prime_i \cap D(s_i)} \varphi(q) \ps{i}(q) \, dq\\
        &+\sum_{i=m+1}^n\int_{\pi_i \cap D(s_i)} \varphi(q) \ps{i}(q) \, dq.
    \end{aligned}
    \end{equation}
\end{lemma}

\begin{proof}
By construction, the first $m$ sensors are mobile, and for each $i \in \Nn_m$, the local coverage region of sensor $s_i$ is $\Pi^\prime_i \cap D(s_i)$. Here, $\Pi^\prime_i$ excludes the coverage of any stationary neighbor $s_j$ ($j>m$) wherever that neighbor’s sensing probability exceeds $\ps{i}(\cdot)$. Hence, the expression: 
\[
\int_{\Pi^\prime_i \,\cap\, D(s_i)} \varphi(q)\,\ps{i}(q)\,dq
\]
captures exactly the portion of the field covered by sensor $s_i$ (and not covered more strongly by a stationary sensor).

For each stationary sensor $s_i$ with $i\in\{m+1,\dots,n\}$, its coverage region within the Voronoi partition $\{\pi_1,\dots,\pi_n\}$ is simply $\pi_i \cap D(s_i)$, since these sensors do not move. Summing these two contributions covers the entire area covered by the network without overlap, leading to~\eqref{eq:fbreak}. Thus, the overall coverage $F$ is a combination of the mobile sensors’ coverage contributions (over $\Pi_i^\prime$) and the stationary sensors’ contributions (over $\pi_i$), yielding~\eqref{eq:fbreak}.
\end{proof}

Moving the mobile sensors cannot improve part of the coverage corresponding to the stationary sensors in~\eqref{eq:fbreak}. Thus, Lemma~\ref{lemma:fbreak} shows that iteratively updating and solving $m$ local coverage maximization problems:
\begin{equation}
    \begin{aligned}
        &\max_{\xs{i}} \quad &&F_i\\
        &\text{subject to} \quad &&\xs{i}\in \Pi_i^\prime\cap\Phi(\xs{i})
    \end{aligned}
\end{equation}
for all $i\in\Nn_m$, maximizes the overall coverage.

\section{Proposed Gradient-Based Deployment Strategy}
\label{sec: strategy}

This section outlines a distributed gradient-based deployment strategy to maximize coverage in a hybrid WSN with probabilistic sensing. The approach leverages the gradient of the local coverage function \( F_i \) as the optimal direction for mobile sensors to enhance their individual contributions to the overall coverage \( F \). This concept, rooted in gradient ascent, has proven effective for heterogeneous and hybrid WSNs with deterministic sensing models~\cite{habibi2017,mosalli2023}.

\begin{figure}
    \centering
    \includegraphics[width=0.75\linewidth]{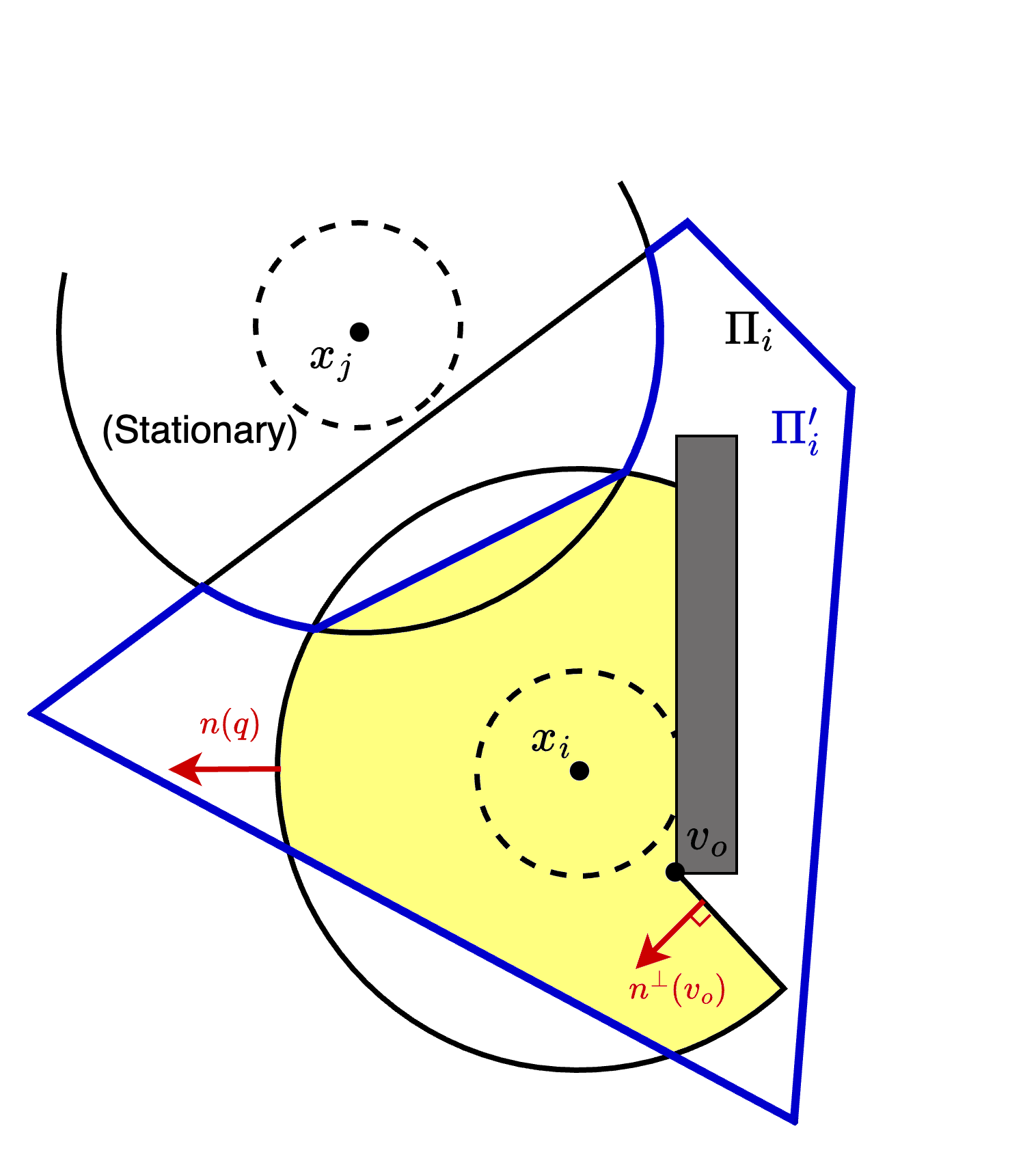}
    \caption{Sample sensor configuration in a local coverage problem}
    \label{fig:vd}
\end{figure}

The gradient of \( F_i \) with respect to \(\xs{i}\) accounts for both the variation in the sensing probability \(\ps{i}(q)\) and changes in the region \(\Pi^\prime_i \cap D(s_i)\). Adopting an approach similar to that in~\cite{mosalli2023}, the boundary of this region is assumed to comprise line segments from the edges of \(\Pi_i\), edges of obstacles facing the sensor, sections excluding stationary neighbors’ sensing ranges per~\eqref{eq:pip}, and parts of \( s_i \)’s sensing range \( D(s_i) \). Additionally, assume that \( P \) obstacle vertices \( v_o^1, v_o^2, \dots, v_o^P \) determine the cut-off edges in the sensor’s visible region \(\Phi(\xs{i})\). A sample case of the local coverage problem is illustrated in Figure~\ref{fig:vd}, where the polygon $\Pi_i$ and region $\Pi_i^\prime$ are shown by black and blue edges, respectively, and the yellow area is the region covered by sensor $s_i$.

\begin{theorem}
\label{thm:grad}
    Consider a hybrid homogeneous WSN in the presence of obstacles, as described earlier. Consider the mobile sensor \( s_i \) (\( i \in \Nn_m \)) with its assigned Voronoi region \(\Pi_i\). Let $R(s_i)$ be part of $D(s_i)$ where $\rs{}^{\min}\leq \| q-\xs{i}\|\leq \rs{}^{\max}$. Let also \(\partial D(s_i)\) denote the portion of the \( D(s_i) \) boundary within \(\Pi^\prime_i\), and \( l_o \) be the radial line segment from \( v_o \) to the perimeter of \( D(s_i) \). Define vectors \( n(q) \) and \( n^\perp(q) \) as the unit vector in the direction of and perpendicular to \( q - \xs{i} \), respectively, for any \( q \in \F \). Then, the gradient of its local coverage function in \eqref{eq:Fi} is derived as:
    \begin{equation}
        \nabla_{\xs{i}} F_i = \nabla_{\xs{i}}^D F_i + \nabla_{\xs{i}}^{\partial D} F_i + \sum_{k=1}^{P} \nabla_{\xs{i}}^{v_o^k} F_i,
    \end{equation}
    where:
    \begin{subequations}
    \label{eq:gradient}
    \begin{align}
    \label{eq:gradient_D}
        &\nabla_{\xs{i}}^D F_i= \alpha \int_{\Pi^\prime_i \cap R(s_i)} n(q) \varphi(q) \ps{i}(q) \, dq, \\
        \label{eq:gradient_dD}
        &\nabla_{\xs{i}}^{\partial D} F_i= \int_{\partial D(s_i)} n(q) \varphi(q) \ps{i}(q) \, dl, \\
        \label{eq:gradient_obs}
        &\nabla_{\xs{i}}^{v_o} F_i = n^\perp(v_o) \int_{l_o} \varphi(q) \ps{i}(q) \, dl.
    \end{align}
    \end{subequations}
\end{theorem}
\begin{proof}
The gradient \(\nabla_{\xs{i}} F_i\) accounts for changes in \(\ps{i}(q)\) and the region \(\Pi^\prime_i \cap D(s_i)\). Using the Leibniz integral rule~\cite{uryasev1995derivatives}:
\begin{equation*}
    \begin{aligned}
        \nabla_{\xs{i}} F_i = &\int_{\Pi^\prime_i \cap D(s_i)} \varphi(q) \nabla_{\xs{i}} \ps{i}(q) \, dq\\
        &+ \int_{\partial (\Pi^\prime_i \cap D(s_i))} \varphi(q) \ps{i}(q) \mathbf{v}(q) n(q) \, dl,
    \end{aligned}
\end{equation*}
where $\mathbf{v}(q)$ is the gradient of the boundary with respect to $\xs{i}$. The boundary term can be broken into two parts: corresponding to the boundary of the sensing ranges and the segments created by obstacles. It is shown in~\cite{habibi2017,mosalli2023} that these terms are equivalent to~\eqref{eq:gradient_dD} and~\eqref{eq:gradient_obs}, respectively.

Moreover, for the first term, which is a surface integral over the covered region by $s_i$, the gradient of the Elfes model given by~\eqref{eq:elfes} can be expressed as:
\[
\nabla_{\xs{i}} \ps{i}(q) = \begin{cases}
    \alpha\ps{i}(q) n(q) \, &,\, \rs{}^{\min}\leq \| q-\xs{i}\|\leq \rs{}^{\max}\\
    0\, &,\, \text{otherwise}
\end{cases},
\]
since:
\[
\nabla_{\xs{i}} \| q - \xs{i} \| = -\frac{q - \xs{i}}{\| q - \xs{i} \|} = -n(q).
\]
Therefore:
\[
\nabla_{\xs{i}}^D F_i = \alpha \int_{\Pi^\prime_i \cap R(s_i)} n(q) \varphi(q) \ps{i}(q) \, dq.
\]
\end{proof}

In Theorem~\ref{thm:grad}, the first term \(\nabla_{\xs{i}}^D F_i\) is a surface integral over the interior of \(\Pi^\prime_i \cap D(s_i)\), similar to the local coverage, while the remaining terms are line integrals along the boundary segments. These integrals can be numerically approximated with prespecified precision using discrete methods, as demonstrated in~\cite{mosalli2023}. The boundary term \(\nabla_{\xs{i}}^{\partial D} F_i\) captures the effect of the shifting sensing range, and the obstacle-related terms \(\nabla_{\xs{i}}^{v_o^k} F_i\) account for visibility constraints imposed by \(\obs\).

To translate this gradient into sensor movement, the authors in~\cite{habibi2017,mosalli2023} employed line search to determine the optimal step size along \(\nabla_{\xs{i}} F_i\). While effective in theory, this method increases computational complexity by requiring multiple evaluations of \( F_i \), overloading resource-limited nodes, and assumes unbounded movement per iteration, neglecting energy and displacement constraints. To address these issues, we propose an approach to set the step size dynamically. The method initially expands the step size and then contracts it over time, facilitating exploration with guaranteed convergence. This is expressed as:
\begin{equation}
\label{eq:step}
\xs{i}(t+1) = \xs{i}(t) + \eta_t \frac{\nabla_{\xs{i}} F_i}{\|\nabla_{\xs{i}} F_i\|},
\end{equation}
where \(\eta_t > 0\) scales the movement based on the iteration count \( t \):
\begin{equation}
\label{eq:dynamic_step}
\eta_t = \min \left( \eta_{\max}, \eta_0 t e^{-\beta t} |\nabla_{\xs{i}} F_i\|\right),
\end{equation}
with \(\eta_0 > 0\) as a base step size and \(\beta > 0\) as a decay parameter, peaking at \( t = \frac{1}{\beta} \).

\begin{remark}
\label{rem:eta_choice}
This formulation increases step sizes early in the deployment (for \( t < \frac{1}{\beta} \)), accelerating sensor repositioning toward critical coverage areas, and then decreases them exponentially (for \( t > \frac{1}{\beta} \)), allowing finer adjustments as the network reaches a steady state. The upper bound \(\eta_{\max}\) ensures movements do not exceed practical limits, addressing energy and connectivity constraints ignored by line search. By utilizing the precomputed \(\nabla_{\xs{i}} F_i\) without requiring additional \( F_i \) evaluations, this approach reduces computational overhead, providing a scalable and efficient solution that adapts to the probabilistic sensing dynamics of hybrid WSNs. However, this choice of \(\eta_t\) is not unique. For instance, an alternative could be \(\eta_t = \min \left( \eta_{\max}, \frac{\eta_0}{\varphi(\xs{i})}\|\nabla_{\xs{i}} F_i\| \right)\), which increases step sizes in low-priority regions to accelerate movement toward critical areas. This alternative is ruled out due to potential numerical complications, as \(\varphi(q)\) may be zero at some points, leading to undefined step sizes that could destabilize the algorithm.
\end{remark}

The candidate position \(\xs{i}(t+1)\) must remain within the feasible region \(\Pi_i^\prime \cap \Phi(\xs{i})\) to ensure the sensor stays in its designated area and travels unobstructed. If the update in~\eqref{eq:step} places \(\xs{i}(t+1)\) outside this region or behind an obstacle, it is mapped to the closest point on the boundary of \(\Pi_i^\prime \cap \Phi(\xs{i})\). This projection maintains the sensor’s assignment to \(\Pi_i^\prime\) and ensures a clear line of sight from \(\xs{i}(t)\) to the new position, avoiding collisions with \(\obs\).

Before moving to the candidate position \(\xs{i}(t+1)\), sensor \( s_i \) evaluates whether the local coverage \( F_i \) with respect to its current Voronoi region \(\Pi_i\) increases if it relocates there. If the increase \( F_i(\xs{i}(t+1)) - F_i(\xs{i}(t)) \) exceeds a predefined threshold \(\epsilon > 0\), the sensor moves to \(\xs{i}(t+1)\); otherwise, it remains at \(\xs{i}(t)\). This ensures movement occurs only when sufficiently beneficial, conserving energy for marginal gains. Furthermore, if \( s_i \) and all its neighbors in \(\N_i\) cease moving at an iteration (i.e., their coverage improvements fall below \(\epsilon\)), \( s_i \) halts further iterations, marking itself as converged. Nevertheless, a sensor that has converged may resume movement if a neighbor starts moving again, due to a neighbor's neighbor relocating. This flexibility allows the network to adapt to global changes effectively. The deployment is terminated, and the entire WSN converges when all sensors reach this state, indicating a steady-state configuration where additional repositioning yields negligible improvement. The deployment strategy is formalized in Algorithm~\ref{alg} for each mobile sensor \( s_i \) (\( i\in\Nn_m\)).
\begin{algorithm}
\caption{Distributed gradient-based deployment for mobile sensors}
\label{alg}
\begin{algorithmic}[1]
    \State \textbf{Init}: \(\xs{i}(0)\), \(\epsilon > 0\), \(\eta_{\max} > 0\), \(t \gets 0\), \(\text{converged}(i) \gets \text{F}\), \(\text{moved}(i) \gets \text{T}\)
    \While{\(\neg \text{all converged}\)}
        \State Get \(\mathcal{N}_i\) per~\eqref{eq:neighbors}, share \(\xs{j}\), \(\text{moved}(j)\), \(j \in \mathcal{N}_i\)
        \If{\(\text{converged}(i) \land \exists j \in \mathcal{N}_i: \text{moved}(j)\)}
            \State \(\text{converged}(i) \gets \text{F}\)
        \EndIf
        \If{\(\neg \text{converged}(i)\)}
            \State Build \(\Pi_i\) from \(\xs{i}(t)\), \(\{\xs{j} \mid j \in \mathcal{N}_i\}\), refine to \(\Pi_i^\prime\) per~\eqref{eq:pip}
            \State Compute \(\nabla_{\xs{i}} F_i\) per~\eqref{eq:gradient}
            \State Get \(\eta_t\) per~\eqref{eq:dynamic_step}, set \(\xs{i}(t+1)\)
            \State Project \(\xs{i}(t+1)\) if outside \(\Pi_i^\prime \cap \Phi(\xs{i})\)
            \State Compute \( F_i(\xs{i}(t+1)) \), \( F_i(\xs{i}(t)) \)
            \If{\( F_i(\xs{i}(t+1)) - F_i(\xs{i}(t)) > \epsilon \)}
                \State \(\text{moved}(i) \gets \text{T}\)
            \Else
                \State \(\xs{i}(t+1) \gets \xs{i}(t)\), \(\text{moved}(i) \gets \text{F}\)
                \If{\(\forall j \in \mathcal{N}_i: \neg \text{moved}(j)\)}
                    \State \(\text{converged}(i) \gets \text{T}\)
                \EndIf
            \EndIf
        \Else
            \State \(\text{moved}(i) \gets \text{F}\)
        \EndIf
        \State \( t \gets t + 1 \)
    \EndWhile
    \State \textbf{Output}: \(\xs{i}(t)\)
\end{algorithmic}
\end{algorithm}
The WSN converges when all sensors satisfy the convergence condition, with stationary sensors fixed throughout.

\begin{remark}
\label{rem:connectivity}
The proposed algorithm ensures network connectivity and global convergence under the assumption that the initial communication graph \(\mathcal{G}\) is connected. Connectivity is maintained throughout the deployment by constraining sensor movements within the communication radius \(\rc{}\). The step size \(\eta_t\) is bounded by \(\eta_{\max}\), and each candidate position \(\xs{i}(t+1)\) is projected onto \(\Pi_i^\prime \cap \Phi(\xs{i})\), ensuring an unobstructed path that preserves existing links with neighbors in \(\N_i\). Consequently, the network remains connected at each iteration. Local convergence occurs when a sensor and all its neighbors cease moving, as their coverage improvements fall below the threshold \(\epsilon\). Given persistent connectivity, these local conditions propagate across the network, leading to global convergence of all sensors to a steady-state configuration.
\end{remark}

\begin{remark}
\label{rem:coverage}
The total coverage \( F \) of the network is non-decreasing and converges in finite time. Each sensor \( s_i \) moves to the candidate point \(\xs{i}(t+1)\) only if the local coverage \( F_i \) increases by at least \(\epsilon > 0\), ensuring that every repositioning enhances the overall coverage~\cite{mahboubi2017}. Since \( F \) is bounded above by the maximum possible coverage over the entire sensing field \(\F\), its non-decreasing behavior guarantees convergence. The minimum improvement threshold \(\epsilon\), on the other hand, prevents excessively small steps. This condition, along with the diminishing step size \(\eta_t\) over iterations, enforces finite-time convergence to a configuration where further movements yield negligible gains.
\end{remark}

\section{Simulation Results}
\label{sec: sim}

In this section, the performance of the proposed strategy is investigated in different scenarios. The parameters for the examples, including \(\beta\), \(\eta_0\), \(\eta_{\max}\), and \(\epsilon\), are determined through trial and error to achieve high coverage performance while ensuring sufficiently fast convergence and adaptability to varying priority functions and field conditions.

\begin{example}
\label{subsec:example1}

This example evaluates the proposed gradient-based deployment strategy in a square sensing field \(\F\) of size 20 with a uniform priority function \(\varphi(q) = 1\) for all \( q \in \F \), and two obstacles. A network of 35 sensors is deployed, with sensing parameters \(\rs{}^{\min} = 0.5\), \(\rs{}^{\max} = 2\), and \(\alpha = 1\) in the Elfes model. Five sensors are stationary and dispersed across the field, while the remaining 30 are mobile, initially clustered in a small region near the center. The communication radius is set as \(\rc{} = 3 \rs{}^{\max}\). The simulation parameters are summarized in Table~\ref{tab:params}.
\begin{table}[t]
\centering
\caption{Simulation Parameters for Examples 1 and 2}
\label{tab:params}
\begin{tabular}{|c|c|c|}
\hline
\textbf{Parameter} & \textbf{Example 1} & \textbf{Example 2} \\
\hline
Priority function \(\varphi(q)\) & Uniform & Gaussian \\
Stationary sensors & 5 & 0 \\
Mobile sensors & 30 & 30 \\
\(\rs{}^{\min}\) & 0.5 & 1 \\
\(\rs{}^{\max}\) & 2 & 2 \\
\(\alpha\) & 1 & 1 \\
\(\rc{}\) & 6 & 6 \\
\(\beta\) & 0.04 & 0.04 \\
\(\epsilon\) & \(10^{-3}\) & \(10^{-3}\) \\
\(\eta_0\) & 0.1 & 0.1 \\
\(\eta_{\max}\) & 2 & 2 \\
\hline
\end{tabular}
\end{table}

\begin{figure}[t]
    \centering
    \includegraphics[width=0.5\textwidth,trim={5cm 1cm 5cm 0cm},clip]{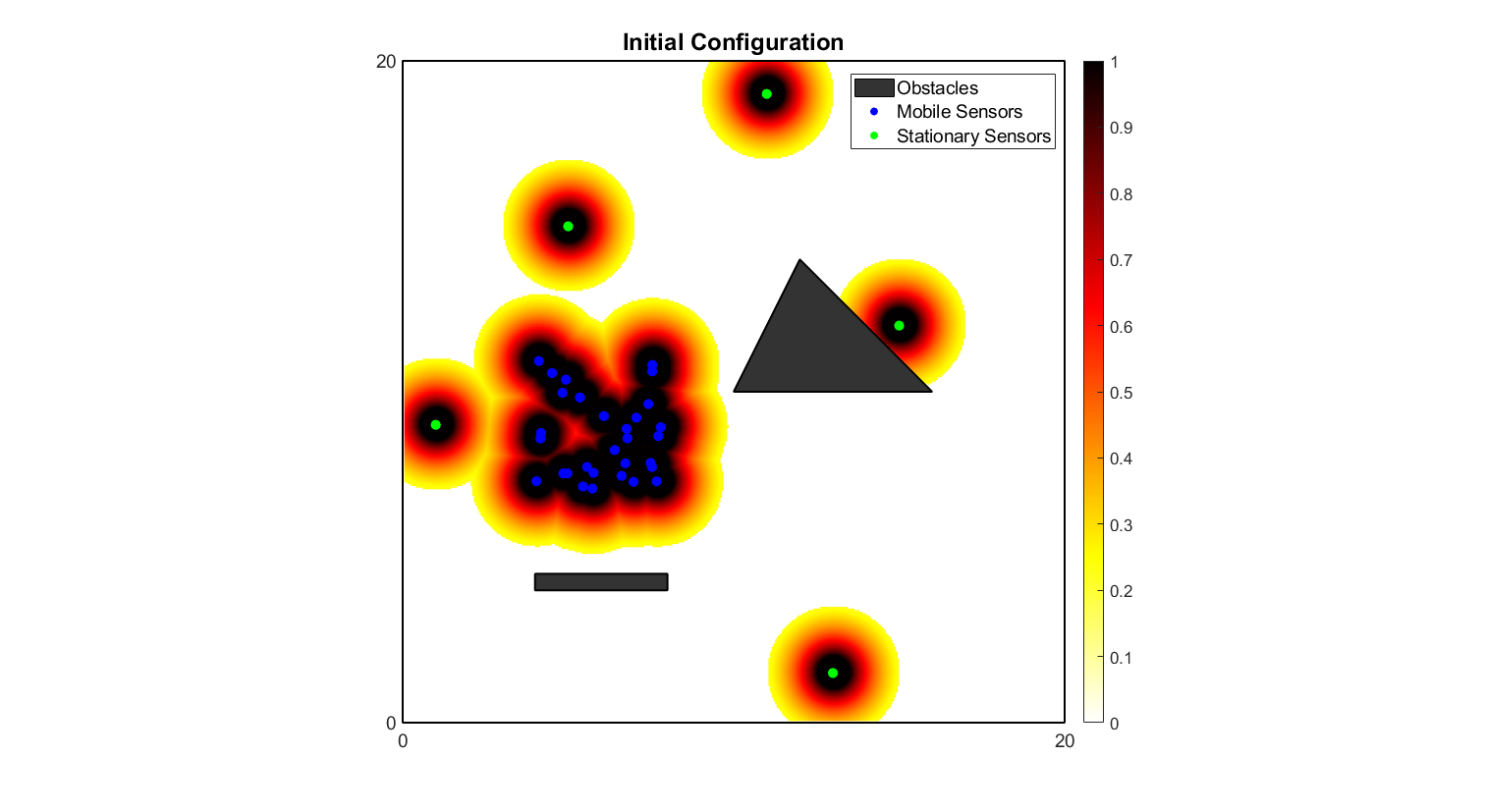}
    \caption{Initial configuration of the WSN in Example 1}
    \label{fig:initial_config}
\end{figure}

\begin{figure}[t]
    \centering
    \includegraphics[width=0.5\textwidth,trim={5cm 1cm 5cm 0cm},clip]{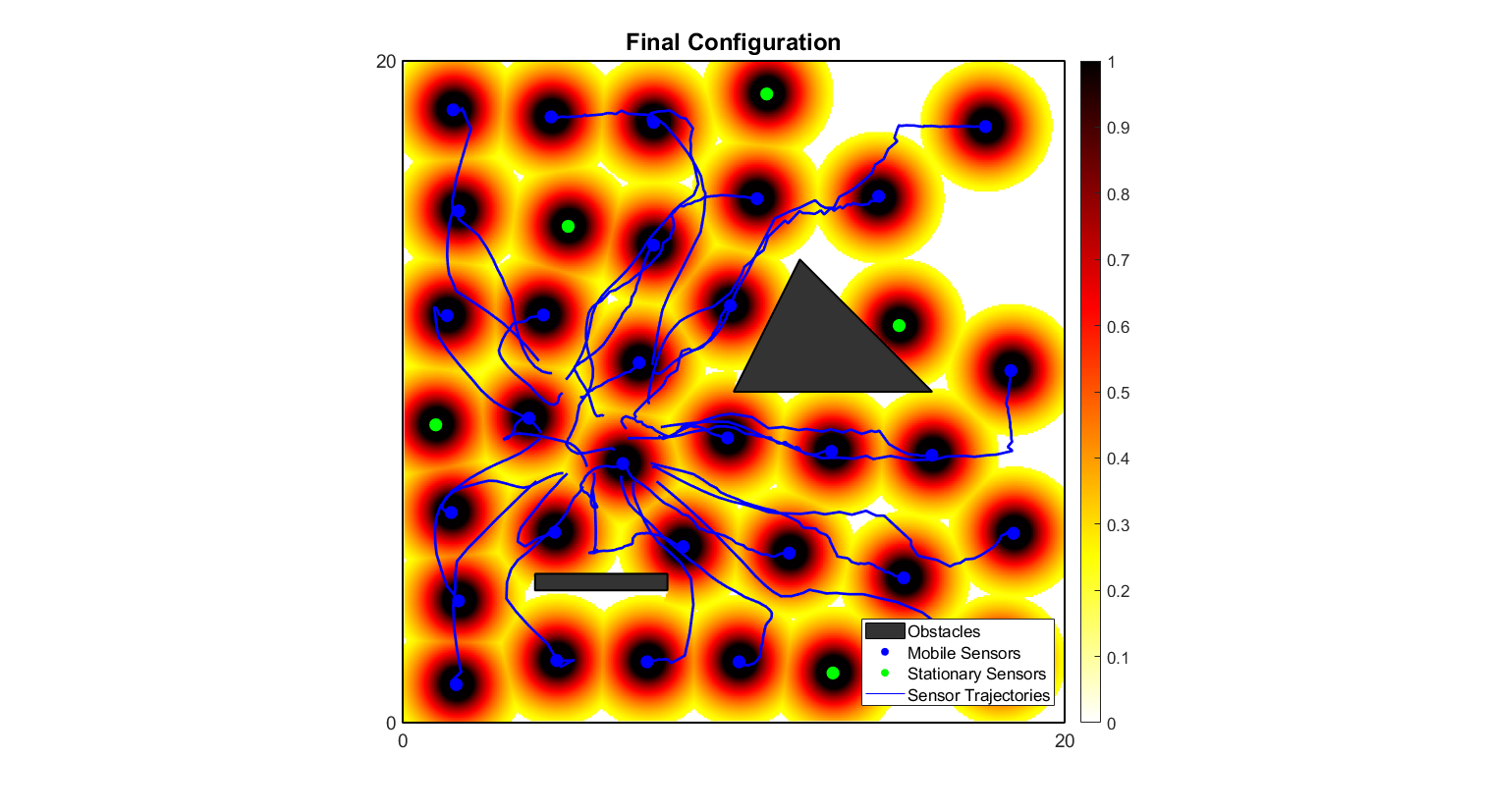}
    \caption{Final configuration of the WSN in Example 1}
    \label{fig:final_config}
\end{figure}

\begin{figure}[t]
    \centering
    \includegraphics[width=0.45\textwidth]{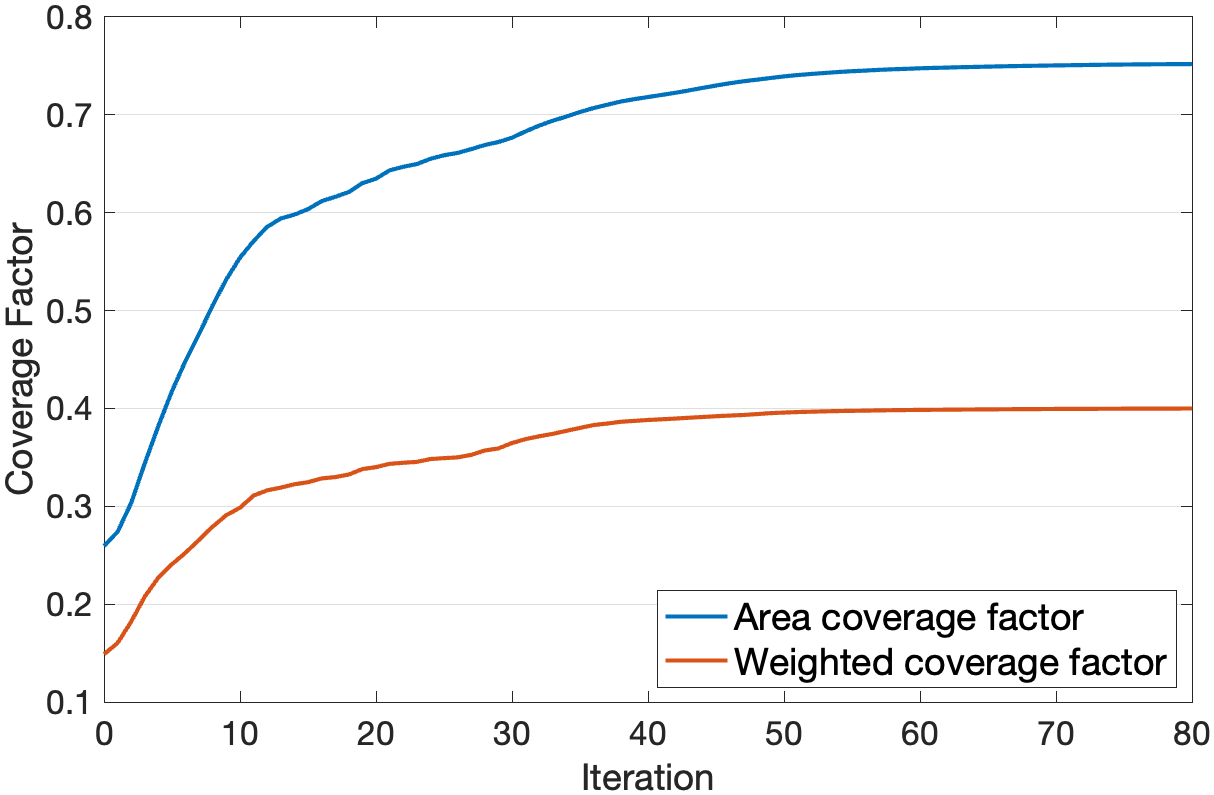}
    \caption{Area coverage factor and weighted coverage factor over 60 iterations, showing significant improvement in both metrics}
    \label{fig:coverage_graph}
\end{figure}

Figures~\ref{fig:initial_config} and~\ref{fig:final_config} illustrate, respectively, the initial configuration and the final one under the proposed approach. The obstacles are depicted as two solid polygons: a triangle and a rectangle. Initially, mobile sensors are concentrated in the center, leading to overlapping sensing ranges and poor coverage of the ROI. After 60 iterations, the mobile sensors have dispersed effectively, navigating around the obstacles while maintaining connectivity within \(\rc{}\). The trajectories (blue lines in Figure~\ref{fig:final_config}) show how sensors adjust positions to cover previously uncovered areas, leveraging the probabilistic Elfes model. Unlike deterministic sensing models, this approach accounts for the gradual decay in detection probability, ensuring a more effective coverage outcome.

Two key performance indicators (KPIs) are defined to quantify coverage. The \textit{area coverage factor} is the ratio of the covered area (where \(\ps{}(q) > 0\)) to the total area of \(\F\). The \textit{weighted coverage factor} is the ratio of the total network weighted coverage to the integral of the priority function over \(\F\), i.e., \(\frac{F}{\int_{\F} \varphi(q) \, dq}\). Since \(\varphi(q) = 1\), the denominator simplifies to the area of \(\F\) (after excluding obstacles). Figure~\ref{fig:coverage_graph} shows that the strategy increases the weighted coverage factor from 15\% to 48\% and the area coverage factor from 27\% to 90\%, demonstrating significant improvement, resulting in most of the ROI being covered by sensors without much overlap in local sensing areas. However, it is observed that although the total covered keeps improving, the total weighted coverage has almost converged after 60 iterations, meaning that the excessive movements do not actually benefit the network coverage. 

The choice of algorithm parameters may significantly impact performance. A high \(\eta_0\) (e.g., \(\eta_0 > 1\)) can cause large jumps early on, potentially overshooting optimal positions and not reaching the coverage increase threshold \(\epsilon\), leading to premature convergence. For instance, if sensor $i, i\in\Nn_m$, moves too far in one iteration, the new position might not satisfy \( F_i(\xs{i}(t+1)) - F_i(\xs{i}(t)) > \epsilon \), causing it to remain stationary thereafter. Similarly, the decay parameter \(\beta\) influences the peak timing and decay rate. A small \(\beta\) delays the peak, prolonging large steps and risking oscillatory behavior, while a large \(\beta\) results in decaying too quickly and limiting exploration. The chosen \(\beta = 0.04\) balances these effects in the above example. The threshold \(\epsilon = 10^{-3}\) ensures finite-time convergence by stopping movements when improvements become negligible, though a smaller \(\epsilon\) might allow finer optimization at the cost of more iterations.
\end{example}

\begin{example}
    \label{subsec:example2}

This example evaluates the proposed strategy in a square sensing field \(\F\) with the same size as before, without obstacles, but with a non-uniform priority function. The priority function \(\varphi(q)\) is defined as the maximum of two Gaussian distributions centered at focal points \([7, 10]\) and \([13, 10]\):
\begin{equation}
\label{eq:priority}
\begin{aligned}
\varphi_1(q) &= \exp\left(-0.02 \left\| q - [7, 10] \right\|^2\right), \\
\varphi_2(q) &= \exp\left(-0.05 \left\| q - [13, 10] \right\|^2\right), \\
\varphi(q) &= \max\left(\varphi_1(q), \varphi_2(q)\right).
\end{aligned}
\end{equation}
A network of 30 mobile sensors, with no stationary sensors, is deployed with sensing parameters \(\rs{}^{\min} = 1\), \(\rs{}^{\max} = 2\), and \(\alpha = 1\) in the Elfes model. The communication radius is \(\rc{} = 3 \rs{}^{\max} = 6\), and the algorithm parameters are the same as Example 1 except for the minimum sensing radius.

\begin{figure}[t]
    \centering
    \includegraphics[width=0.4\textwidth,trim={5cm 0pt 5cm 0pt},clip]{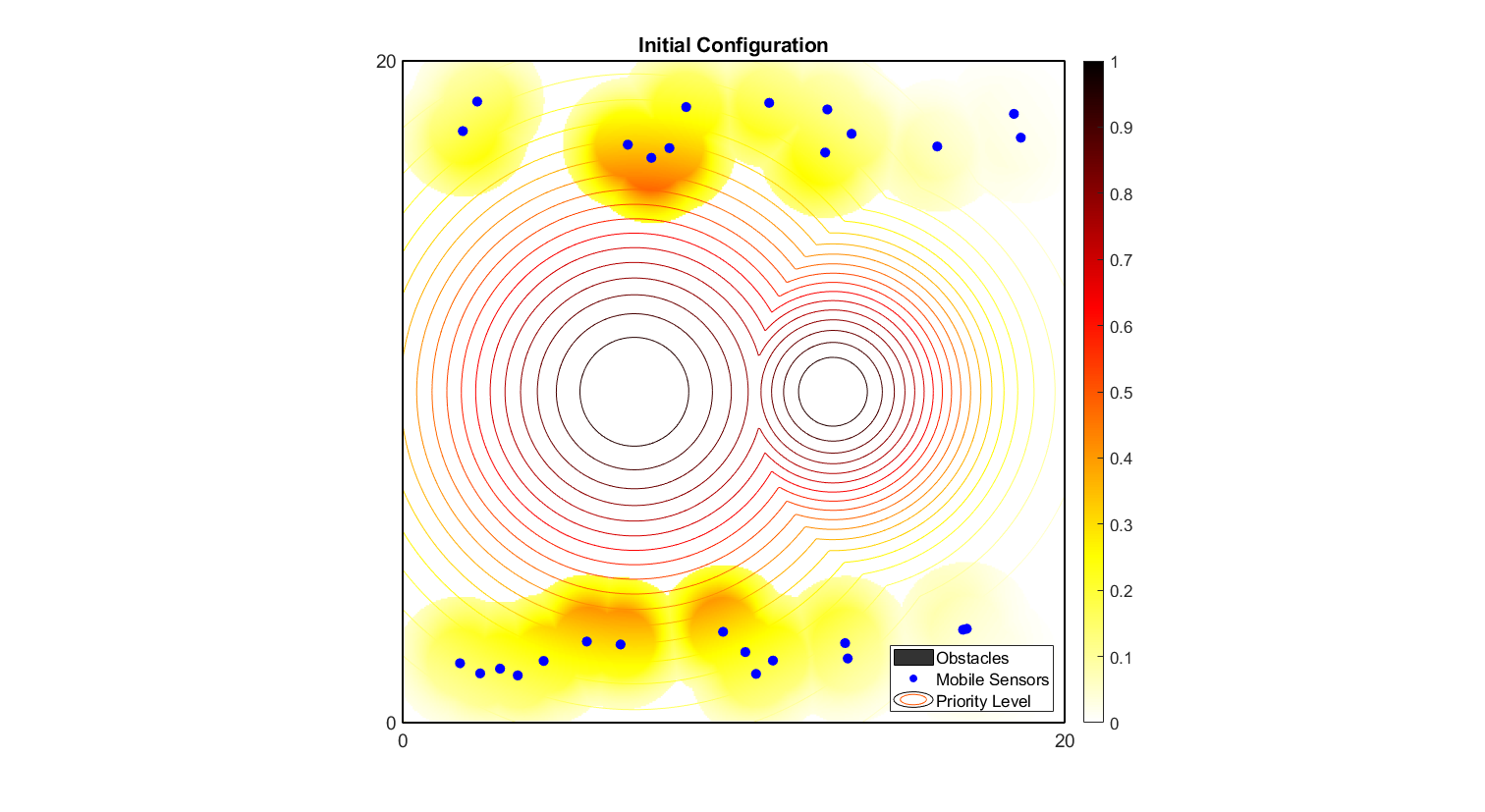}
    \caption{Quality of sensing in the initial configuration of the WSN in Example 2.}
    \label{fig:initial_config_ex2}
\end{figure}

\begin{figure}[t]
    \centering
    \includegraphics[width=0.4\textwidth,trim={5cm 0pt 5cm 0pt},clip]{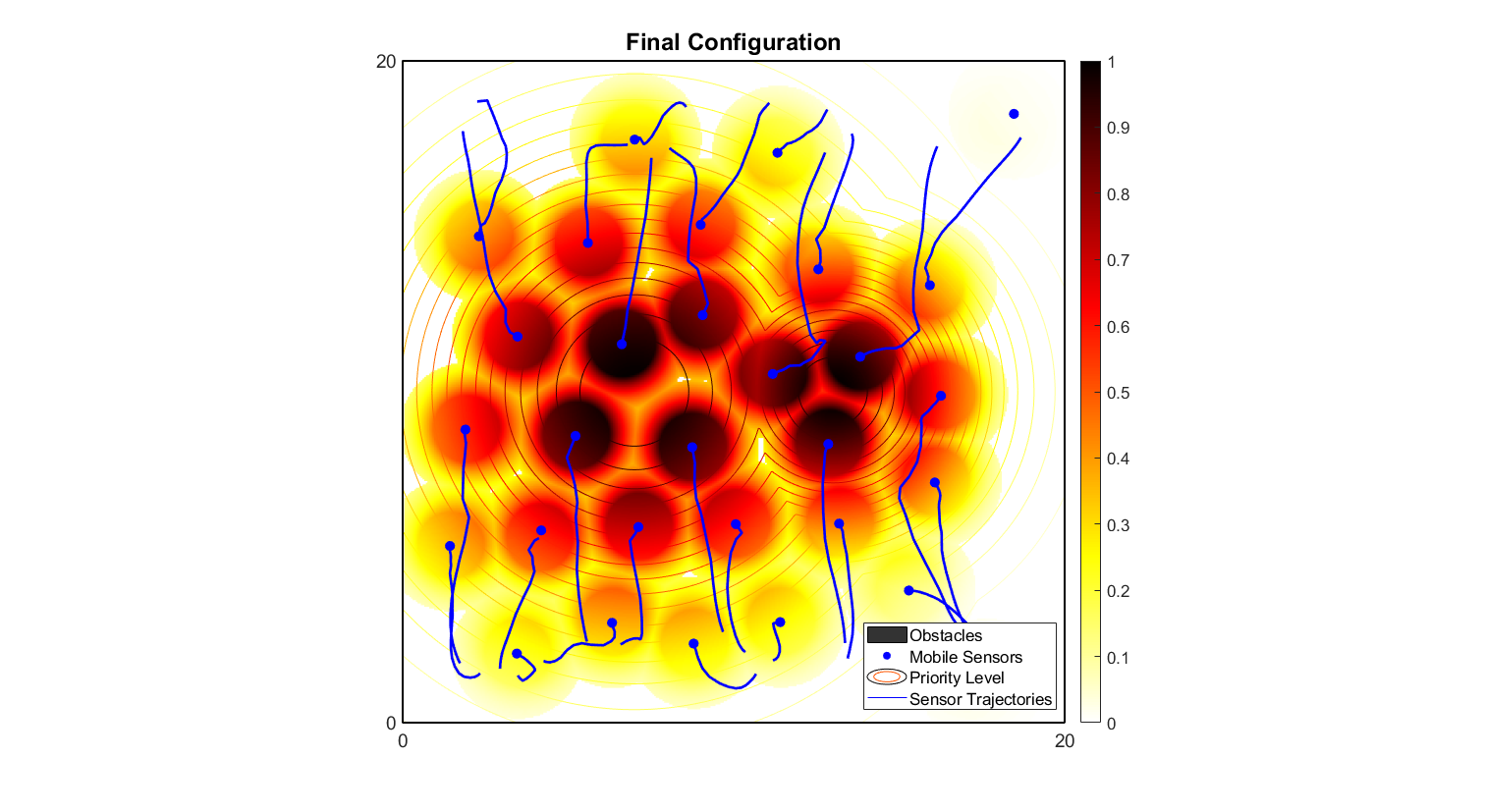}
    \caption{Quality of sensing in the final configuration of the WSN in Example 2.}
    \label{fig:final_config_ex2}
\end{figure}

Figures~\ref{fig:initial_config_ex2} and~\ref{fig:final_config_ex2} depict the initial and final configurations, respectively. The priority function \(\varphi(q)\) is visualized through contour curves, with higher values near the focal points \([7, 10]\) and \([13, 10]\). The sensing quality defined as \(\ps{}(q) \varphi(q)\) is represented by color shades, where darker shades indicate higher quality. Initially, the mobile sensors are positioned at the top (near \( y = 20 \)) and bottom (near \( y = 0 \)) of the field, far from the high-priority regions centered at \( y = 10 \). This results in low sensing quality across the central high-priority areas, as seen in the sparse color distribution in Figure~\ref{fig:initial_config_ex2}. Under the proposed algorithm, the sensors converge toward the high-priority regions, clustering around the two focal points, as shown in Figure~\ref{fig:final_config_ex2}. The trajectories illustrate how sensors directly move inward in the absence of obstacles, effectively covering the areas with the highest \(\varphi(q)\).

\section{Conclusions}
\label{sec: conc}

This paper introduced a distributed gradient-based deployment strategy for hybrid homogeneous WSNs, addressing the challenges of probabilistic sensing and environmental obstacles. By reformulating coverage, introducing local terms via Voronoi partitioning, and employing the Elfes model, the approach enables mobile sensors to enhance coverage using local gradients, while a dynamic step size and threshold-based movement ensure energy efficiency and desired coverage performance. The strategy’s distributed nature and its ability to handle obstacles through visibility adjustments make it scalable for large networks. Simulation results validate the algorithm's effectiveness, showing significant coverage improvements over static configurations in two examples. Future work could explore adaptive priority functions for dynamic environments, incorporate heterogeneous sensor parameters, and integrate machine learning for real-time optimization, further enhancing applicability to critical monitoring tasks.
\end{example}

\bibliographystyle{ieeetr}
\bibliography{references}

\end{document}